\documentclass[runningheads]{llncs}
\usepackage{float}
\usepackage{graphicx}
\usepackage{longtable}
\usepackage{booktabs}
\usepackage{multirow}
\usepackage{hyperref}
\usepackage{amsmath}
\usepackage{csquotes}

\usepackage{cite}
\begin{document}
 \title{An Ensemble Approach for Brain Tumor Segmentation and Synthesis}

%
%
\author{
Juampablo E. Heras Rivera\inst{1*} \and
Agamdeep S. Chopra\inst{1*} \and
Tianyi Ren\inst{1*} \and
Hitender Oswal \inst{2} \and
Yutong Pan\inst{2} \and
Zineb Sordo\inst{3} \and
Sophie Walters\inst{4} \and
William Henry\inst{5} \and
Hooman Mohammadi\inst{1} \and
Riley Olson\inst{6} \and
Fargol Rezayaraghi\inst{1}\and
Tyson Lam\inst{1} \and
Akshay Jaikanth\inst{1} \and
Pavan Kancharla\inst{1} \and
Jacob Ruzevick\inst{7}\and
Daniela Ushizima\inst{3} \and
Mehmet Kurt\inst{1} }

%
\authorrunning{J.E. Heras Rivera et al.}
%
\institute{Department of Mechanical Engineering, University of Washington, Seattle, WA, USA \and Paul G. Allen School of Computer Science, University of Washington, Seattle, WA, USA \and
Lawrence Berkeley National Laboratory, Berkeley, CA, USA \and
Department of Bioengineering, University of Washington, Seattle, WA, USA \and 
Department of Electrical and Computer Engineering, University of Washington, Seattle, WA\and
Watauga High School, Boone, NC, USA \and
Department of Neurological Surgery, University of Washington, Seattle, WA, USA\\
* authors contributed equally to this paper
}

\maketitle              
\begin{abstract}
The integration of machine learning in magnetic resonance imaging (MRI), specifically in neuroimaging, is proving to be incredibly effective, leading to better diagnostic accuracy, accelerated image analysis, and data-driven insights, which can potentially transform patient care.  Deep learning models utilize multiple layers of processing to capture intricate details of complex data, which can then be used on a variety of tasks, including brain tumor classification, segmentation, image synthesis, and registration. Previous research demonstrates high accuracy in tumor segmentation using various model architectures, including nn-UNet and Swin-UNet. U-Mamba, which uses state space modeling, also achieves high accuracy in medical image segmentation.
To leverage these models, we propose a deep learning framework that ensembles these state-of-the-art architectures to achieve accurate segmentation and produce finely synthesized images.


\end{abstract}
\section{Introduction}
Gliomas are the most common primary brain tumors and grow from glia, the nonneuronal support cells in the brain and nervous system. High-grade gliomas are associated with poor prognosis, with a 5-year survival rate of 5-10\% for glioblastoma patients, and a mean survival time of 9 months \cite{GlioblastomaSurvival}. Typical treatment of gliomas generally involves surgical resection and radiation, so accurate localization of the tumor mass is critical for improving patient prognosis.   

Currently, localization of gliomas is done by manual analysis of magnetic resonance imaging (MRI) scans by experienced radiologists. However, due to the heterogenity in appearance and shape of gliomas, manual tumor boundary segmentations are prone to large variability among radiologists of different experience levels \cite{RaterVariability}. These issues necessitate the development of an automated segmentation tool that can leverage insights from MRI data to improve reliability. Recent advancements in deep learning methods have emerged as a possible solution; however, these methods require large amounts of annotated MRI data for training. To this end, the 2024 Brain Tumor Segmentation (BraTS) challenge \cite{BraTS2024} provides a community standard and benchmark for state-of-the-art automated segmentation models based on the largest expert-annotated glioma MRI dataset. 

The BraTS challenge includes multiple sub-challenges; for this year's submission, we participated in the following: \textbf{Adult Glioma Post Treatment}, \textbf{BraTS-Africa}, \textbf{Generalizability}, and \textbf{Synthesis} through local inpainting. All of the considered challenges provide multi-parametric MRI (mpMRI) scans of brain tumor patients, and have been annotated manually by one to four raters, following the same annotation protocol, and their annotations were approved by experienced neuroradiologists \cite{BraTS2024}. 

Our approach to the 2024 BraTS challenge is heavily inspired by the previous year's BraTS winners \cite{FakingIt}, who combined "unconventional" mechanisms for data augmentation and an ensembling approach of 3 state-of-the-art tumor segmentation models to improve model performance. Similarly, our pipeline includes data augmentation using a series of augmentation methods, an ensemble of checkpoints produced by 4 state-of-the-art complementary UNet \cite{3DUNet} based architectures, and an adaptive post-processing step.


\section{Methods}

\subsection{Data}
\subsubsection{BraTS Adult Glioma Post Treatment (AGPT):} This dataset includes scans of glioma patients and corresponding annotations of the non-enhancing necrotic tumor core (NETC), the surrounding non-enhancing FLAIR hyperintensity (SNFH), the enhancing tissue (ET), and a new label for the post-treatement resection cavity (RC), as described in \cite{BraTS2024}. This dataset includes \textbf{2200} annotated training samples.

\subsubsection{BraTS-Africa Challenge dataset:}
This dataset includes 1.5T magnetic field strength MRI scans of glioma patients in Sub-Saharan Africa (SSA), along with annotations of the NETC, SNFH, and ET tumor regions. Due to the limited accessibility to imaging centers in SSA, this dataset is much smaller than the AGPT dataset, consisting of \textbf{60} annotated training samples.

\subsubsection{BraTS Generalizability Across Tumors (BraTS-GoAT):}This dataset includes scans for various brain tumors and patient populations, along with annotations of the NETC, SNFH, and ET regions. The aim of this challenge is to assess the algorithmic generalizability of segmentation models to lesion type, scanning institution, and patient demographics. The driving hypothesis behind this challenge is that a method capable of performing well on multiple segmentation tasks will generalize well to unseen tasks \cite{BraTS2024}. Training data for this challenge consists of preoperative data collected from the BraTS 2023 challenge, including: the full Adult Glioma dataset, the Meningioma dataset with ground truth only available for a subset of the data, and the Brain Metastases data with ground truth only available for a subset of the data. The validation data includes \textit{validation} data from the BraTS 2023 challenge, specifically: the Adult Glioma, BraTs-Africa, Meningiomas, Metastases, and Pediatric Tumors validation datasets. The training set consists of \textbf{1351} samples.

\subsubsection{BraTS Inpainting:}

This dataset includes a collection of masked T1 MRI scans with the corresponding mask. The training data also includes the ground truth image as well as masks covering both healthy and unhealthy regions. Validation and testing data only includes the masked volume and the mask. The baseline is exclusively sourced from the multi-modal BraTS glioma segmentation challenge, with the provided masks generated randomly on the fly. Training and validation sets consist of \textbf{1251} and \textbf{219} samples respectively. The validation data from 2023 BraTS Glioma was used to quantitatively evaluate model performance using randomly generated masks.



\subsection{Evaluation Metrics}

Model performance for the 2024 BraTS Challenge is evaluated using the lesion-wise Dice \cite{zhang2021all, 3DUNet} and Hausdorff Distance 95\% (HD95) metrics \cite{billet2008use} between each region of interest in the ground truth and the corresponding prediction. Lesion-wise metrics prevent the evaluation from favoring models that only detect larger lesions by evaluating on a lesion-by-lesion basis and giving equal weight to each lesion regardless of size \cite{zhang2021all}. Additionally, there are additional penalties for false negative (FN) and false positive (FP) predictions as described in \cite{BraTS2024}. For the inpainting task, metrics like Structural Similarity Index Measure (SSIM) \cite{armanious2019adversarial}, Peak Signal to Noise Ratio (PSNR) \cite{armanious2019adversarial, kang2021deep}, and Mean Square Error (MSE) \cite{almansour2021high} are employed to evaluate image fidelity, quality, and realism. SSIM assesses perceptual similarity by comparing luminance, contrast, and structure, aligning well with human visual perception. PSNR measures the ratio between the maximum possible pixel value and the noise level introduced during inpainting, with higher values indicating better image fidelity. MSE calculates the average squared difference between the inpainted and original pixels, providing a measure of pixel-level accuracy. Together, these metrics offer a comprehensive evaluation of the quality of the synthesized image.



\subsection{Model Architectures}
\subsubsection{Optimized U-Net}\label{section:OptimizedUNet}
We used an adaptation \cite{ren2024optimization} of the optimized U-Net  \cite{futrega2021optimized} architecture used by the winners of the 2021 BraTS challenge as our baseline model architecture for comparison purposes. U-Net has a symmetric U-shape that characterizes architecture and can be divided into two parts, i.e., encoder and decoder. The encoder comprises 5 levels of same-resolution convolutional layers with strided convolution downsampling. The decoder follows the same structure with transpose convolution upsampling and convolution operating on concatenated skip features from the encoder branch at the same level.

\paragraph{Training details}
For the BraTS Adult Glioma Post Treatment and BraTS GoAT challenges, where training data surpassed 1000 samples, 5-fold cross validation was used. Due to the small size of the training dataset, 10-fold cross validation was used for BraTS-Africa.

To train the optimized U-Net framework described in \ref{section:OptimizedUNet}, we use the Adam \cite{AdamOptimizer} optimizer with an exponentially-decaying learning rate starting at $\alpha_0=6e-5$, and subsequent learning rates calculated as $ \alpha_i = \alpha_0 \times \left(1 - \frac{\text{epoch}_i}{\text{epoch}_N} \right)^{0.75}$, where $\alpha_i$ is the learning rate at epoch \(i \in 1, \ldots, N\). Each fold is trained for 100 epochs, which we found to produce acceptable Dice and HD95 scores through experimentation. 
For the rest of the models that were used for model ensemble, the detailed hyperparameters can be found from the corresponding literature.

\subsubsection{RhizoNet}
RhizoNet is a deep-learning based 2D-image segmentation pipeline of plant roots \cite{Sordo2024}, used to automate the process of root image analysis. This pipeline consists of a 2D Residual U-Net taking as input pre-processed small size patches. Given the BraTS-Africa dataset, RhizoNet was modified to a 3D Residual U-Net and the inputs to 3D volume patches with 4 possible classes including background.

The Residual U-Net combines the concepts of residual networks (ResNets) and the U-Net architecture such that each layer in the encoder and decoder is implemented as a residual block which contains skip connections that allow the network to learn residual information: each block consists of a residual unit with 2 subunits. Each subunit has a batch normalization, rectified linear unit (ReLU) activation function, and convolution layers with $3 \times 3$ kernel filters for each convolution and 30\% dropout. The loss function used is the weighted cross-entropy loss.

\subsubsection{nn-UNet}
The state of art model in medical image segmentation model is nnU-Net.  A deep learning-based segmentation method that automatically configures itself including preprocessing, network architecture, training, and post-processing for any new task in the biomedical domain\cite{isensee2021nnu}. It also offers various residual encoder UNets, which improves segmentation performance\cite{isensee2024nnunetrevisited}.

\subsubsection{Swin-UNetR}
Vision Transformers (ViTs) have ushered in a transformative shift in computer vision and medical image analysis \cite{xiao2023transformers}. These models exhibit proficiency in capturing both global and local information through their layered architecture. Additionally, ViTs offer significant scalability, making them highly effective for large-scale training scenarios. Their ability to leverage self-attention mechanisms enables them to integrate and synthesize complex features across diverse image domains, setting a new benchmark in performance and adaptability within the field. Swin UNETR comprises a Swin Transformer encoder that directly utilizes 3D patches and is connected to a CNNbased decoder via skip connections at different resolutions\cite{hatamizadeh2021swin}.

\subsubsection{U-Mamba}

State Space Models (SSMs) has been a growing comptetitor to attention-based transformers, particularly on tasks with complex, long range sequences. With a slew of papers using the novel Mamba architecture, SSMs execute much faster than traditional transformer models and perform better than all traditional models. The U-Mamba architecture is the latest advance for Deep Learning models in the Biomedical imaging domain. The U-Mamba utilizes an encoder-decoder architecture with Mamba blocks and performs on-par--if not better--than current models with reduced computational complexity\cite{U-Mamba}.

\subsubsection{Re-DiffiNet}
In a previous research, we proposed a tumor segmentation framework Re-DiffiNet, which uses diffusion models to refine and improve predictions of a tumor segmentation model (like optimized U-Net) \cite{ren2024re}. We found that while using diffusion models to directly generate tumor masks did lead to improvements in performance over the baseline U-Net, it was the use of discrepancy modeling i.e. predicting the differences between ground truth masks and baseline U-Net's outputs, that led to most significant improvements\cite{ren2024re}.

\subsubsection{DANN}
For the BraTS-Africa Challenge, we apply the Domain Adversarial Neural Network (DANN) \cite{DANN} model architecture to promote transfer learning between the Brats 2023 Adult Glioma dataset and the BraTS-Africa dataset. DANN aims to learn features that are both discriminative and domain-invariant by optimizing a neural network comprising two classifiers: a label predictor, which provides segmentations and is used both during training and testing, and a domain classifier that discriminates between domains during training. The optimization process minimizes the label classifier loss while maximizing the domain classifier loss, promoting the emergence of domain-invariant features through an adversarial optimization. Our implementation of the DANN approach consists of the Optimized U-Net \cite{ren2024optimization} as a backbone, with an additional domain classifier at the bottleneck layer.  In the context of transfer learning between the Brats 2023 Adult Glioma dataset and the BraTS-Africa dataset, we expect the model to leverage domain-invariant features common across domains for segmentation.
In addition to this configuration of DANN, we also used a DANN architecture without the gradient reversal layer, as proposed in \cite{HerasRiveraMIDL}, since this architecture produced improved results for the SNFH (formerly ED) region.


\subsubsection{Attention-UNet}
We used a modified Attention U-Net as the baseline for the inpainting task \cite{oktay2018aunet}. This model integrates attention gates into the skip-connections of a standard U-Net, enhancing image reconstruction by focusing on relevant features and improving contextual understanding.
The model was trained with two different approaches. The first used a supervised strategy combining L1 loss, SSIM loss, Canny Edge loss\cite{abderezaei20223d}, and mask-focused MSE loss. The second utilized a conditional GAN framework \cite{goodfellow2014gans}\cite{isola2018cgan} with the previous supervised loss function, a Wasserstein objective \cite{arjovsky2017wgan}, and gradient penalty \cite{gulrajani2017iwgans}. The models were trained for 1500 epochs, with hyperparameters set based on prior work \cite{isola2018cgan}\cite{arjovsky2017wgan} and empirical testing. The best iteration was selected based on validation metrics and reported in Table \ref{tab:synthesis_model_metrics}.

\begin{figure}[H]
    \centering
    \includegraphics[width=0.7\linewidth]{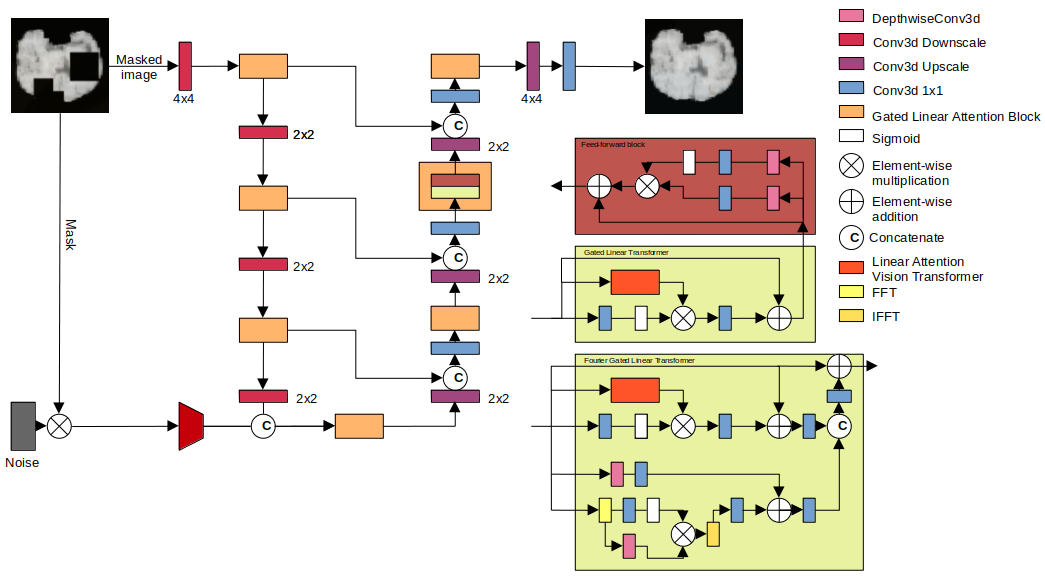}
    \caption{Architecture of our proposed Mask Aware 3D T-Former network for inpainting tasks. There are 2 flavours of the Transformer block, a standard gated linear attention block, and a gated linear attention block with an additional gated Fourier block for enhanced global contextual learning.}
    \label{fig:MA3TF}
\end{figure}

\subsubsection{MA3T-former}
We propose a novel architecture inspired by recent advancements in inpainting and synthesis tasks, leveraging improved attention mechanisms and Fourier transformation-based architectures. Our model extends the T-former model into a 3D domain \cite{Deng2022Tformer} that uses Taylor expansion to derive a more efficient implementation of the attention algorithm. Our enhancements include a mask embedding layer combined with noise inpainting to preserve the stochastic nature of the model. We also developed a second variant that integrates a Fourier Neural Block \cite{li2021fourierneuraloperator} to enhance global contextual understanding. This operator features a modified, learnable self-gating layer, functioning as an active filter to further refine contextual learning. The detailed structure of this model is illustrated in Figure \ref{fig:MA3TF}.
Training was conducted using the WGANs-based optimization framework, similar to that used for the Attention UNet model. We augmented the training process by incorporating the GrokFast algorithm \cite{lee2024grokfast}, which accelerates the delayed generalization phenomenon by amplifying the slow-varying components of gradients. These optimizations, alongside an efficient linear attention implementation, enabled our model to reach 2.35 billion parameters while maintaining low latency during inference. This was achieved with minimal computational resources compared to traditional approaches.

\subsubsection{Data augmentation}

Data augmentation is the method of artificially expanding a limited dataset by applying a variety of transformations to the images during the training phase. Inspired by the winners of the BraTS 2021 Challenge \cite{FakingIt}, we employ data augmentation strategies to reduce the risk of model over-fitting and enhance robustness against MRI artifacts present in the dataset. Basic augmentation strategies include applying Gibbs noise, Gaussian noise, and Gaussian smoothing, scaling intensity, and flipping the image. Novel methods included
\begin{itemize}
    \item \textbf{Motion}: With a probability of 0.1, random MRI motion artifacts are applied to images with degrees and translation of ten with linear image interpolation and one simulated movement. Motion artifacts, such as smearing or ghosting, appear in MRI scans due to subjects physically moving during image acquisition.
    
    \item \textbf{Spike}: With a probability of 0.1, it applies random MRI spike artifacts with intensity sampled uniformly from (1, 3). Spike artifacts, or Herringbone artifacts, appear in MRI scans as stripes in different directions because of spikes in the k-spaces.

    \item\textbf{Bias Field}: With a probability of 0.1, it applies random MRI bias field artifacts to the images with a coefficient of 0.5. In MRI, a bias field is a gradual, low-frequency variation in image intensity caused by inconsistencies in the magnetic field, RF coils, or the patient’s anatomy. 

    \item \textbf{Elastic Deformation}: With a probability of 0.1, random dense elastic deformations are applied to the images with seven control points and max voxel displacement. 

    \item \textbf{Anisotropy}: With a probability of 0.1, the transform randomly scales voxel dimensions along different axes to simulate anisotropic effects with linear image interpolation and downsampling uniformly sampled from (1, 2). 

\end{itemize}

Spike, Bias Field, and Anisotropy were specifically implemented because MRI scans that resulted in low Dice scores presented these effects. 

\subsubsection{Loss function}

The total loss is a weighted sum of  dice and focal, loss. Each of these scores were considered and implemented in the Kurtlab Group BraTS2023 paper \cite{ren2024re}. Additionally, to produce a final loss with better geometric representation, we also included a bounding-box loss and a rotational inertia loss in that sum, inspired by the recently released BiomedParse preview \cite{BiomedParse}. 





\subsubsection{Postprocessing}
After completing ensembling, we implemented a dust removal and volume-based post processing strategy. We first perform dust removal for connected components smaller than 50 voxels for ET, TC, and WT, and relabel voxels based on their surrounding labels\cite{ren2024optimization}. We then calculate the voxel-wise volumes of ET, SNFH, and WT, and find the ET/WT and SNFH/WT ratios. For cases where ET/WT is less than 0.03, we relabel ET voxels with NETC. For cases where SNFH/WT is 1, we relabel SNFH voxels with NETC.  

\section{Results}
\subsection{BraTS-Goat}

The test dataset results in Table 1 indicate that nnU-Net and UMamba models outperform Optimized U-Net and Swin-UNETR across most evaluated metrics. Among all the models, nnU-Net consistently demonstrates superior performance in terms of Dice scores and Leasion-Wise scores, while UMamba shows exceptional performance in reducing HD95 and Lesion-Wise HD95 scores across all regions. The Optimized U-Net model, while showing balanced performance, is generally surpassed by nnU-Net and UMamba. Swin-UNETR lags behind the other models, particularly in HD95 and Lesion-Wise HD95 for the Whole Tumor sege region.
\begin{table}[H]
\centering
\resizebox{\linewidth}{!}{
\begin{tabular}{|c|ccc|ccc|ccc|ccc|}
\hline
\multirow{2}{*}{Model Architecture} & \multicolumn{3}{c|}{Dice [\%]} & \multicolumn{3}{c|}{HD95 [mm]} & \multicolumn{3}{c|}{LW Dice [\%]} & \multicolumn{3}{c|}{LW HD95 [mm]} \\
\cline{2-13}
& WT & ET & TC & WT & ET & TC & WT & ET & TC & WT & ET & TC \\
\hline
Optimized U-Net & 81.44 & 72.25& 75.81& 22.30&  48.47& 29.94& 68.50& 67.85& 71.27& 84.84 & 77.05& 51.79\\
nnU-Net & 88.11 & 79.98 & 82.51 & 17.49 & 37.00 & 23.79 & 82.56 & 78.07 & 79.81 & 39.62 & 48.26 & 34.54 \\
Swin-UNetR & 78.59 & 72.51 & 74.75 & 22.85 & 48.88 & 31.28 & 60.96 & 65.75 & 68.03 & 116.09 & 90.51 & 67.75 \\
UMamba & 85.69 & 77.41 & 82.07 & 14.51 & 37.48 & 18.90 & 78.60 & 73.95 & 78.55 & 50.24 & 62.11 & 38.17 \\
\hline
\end{tabular}
}
\caption{Test performance metrics for the different training pipelines considered. Metrics include Dice, HD95, lesion-wise Dice (LW Dice), and lesion-wise HD95 for whole tumor (WT), tumor core (TC) and ET.}
\label{table:unet_metrics_goat}
\end{table}

The experiment results in Table 2 indicate that applying data augmentation techniques to the Optimized U-Net model with a probability of ten percent will produce the highest Dice scores and the lowest HD95 scores, the only exception being twenty percent data augmentation producing the highest WT Dice score. These results correspond with general augmentation sensitivity; Insufficient augmentation may fail to enhance the model's generalization abilities adequately, whereas excessive augmentation can degrade model performance by training it on data that diverges too far from the original dataset. Comparing Table 1 and Table 2 results of Optimized U-Net, data augmentation had significantly lower HD95 scores and slightly higher Dice scores. 

\begin{table}[H]
\centering
\begin{tabular}{|c|l|c|c|c|c|c|c|c|c|c|}
\hline
\multirow{2}{*}{Fold \#} & \multirow{2}{*}{Probability} & \multicolumn{4}{c|}{Dice [\%]} & \multicolumn{4}{c|}{HD95 [mm]} \\ \cline{3-10} 
 &  & WT & ET & TC & Avg & WT & ET & TC & Avg \\ \hline
\multirow{4}{*}{fold 1} & \textbf{ 5\%} & 89.43 & 86.36 & 81.43 & 85.91 & 8.13 & 5.97 & 4.98 & 6.36 \\ \cline{2-10} 
 & \textbf{10\%} & 89.71 & 86.51 & 82.32 & 86.18 & 6.72 & 5.36 & 4.59 & 5.56 \\ \cline{2-10} 
 & \textbf{20\%} & 89.73 & 86.49 & 81.72 & 85.98 & 6.41 & 5.56 & 4.57 & 5.52 \\ \cline{2-10} 
 & \textbf{30\%} & 89.88 & 85.78 & 80.87 & 85.51 & 7.36 & 6.16 & 5.51 & 6.34 \\ \hline
\multirow{4}{*}{fold 2} & \textbf{5\%} & 89.27 & 84.55 & 79.80 & 84.54 & 7.40 & 6.86 & 5.39 & 6.55 \\ \cline{2-10} 
 & \textbf{10\% }& 89.68 & 84.25 & 79.64 & 84.52 & 7.80 & 6.62 & 5.75 & 6.72 \\ \cline{2-10} 
 & \textbf{20\% }& 89.22 & 83.53 & 78.85 & 83.87 & 7.59 & 6.59 & 5.54 & 6.57 \\ \cline{2-10} 
 & \textbf{30\%} & 89.62 & 84.75 & 79.15 & 84.51 & 7.41 & 6.20 & 5.24 & 6.28 \\ \hline
\multirow{4}{*}{fold 3} & \textbf{5\%} & 90.50 & 86.86 & 83.20 & 86.86 & 7.31 & 5.32 & 4.31 & 5.65 \\ \cline{2-10} 
 & \textbf{10\%} & 90.62 & \textbf{87.71} & 83.37 & \textbf{87.23} & \textbf{5.62} & 6.09 & 4.85 & 5.52 \\ \cline{2-10} 
 & \textbf{20\%} & 90.25 & 86.91 & 82.57 & 86.58 & 7.08 & 6.96 & 5.99 & 5.99 \\ \cline{2-10} 
 & \textbf{30\%} & 90.79 & 86.97 & 81.99 & 86.58 & 6.05 & 6.88 & 5.62 & 6.19 \\ \hline
\multirow{4}{*}{fold 4} & \textbf{5\%} & 87.20 & 83.68 & 78.39 & 83.09 & 8.08 & 6.06 & 5.21 & 6.45 \\ \cline{2-10} 
 & \textbf{10\%} & 86.73 & 82.74 & 78.67 & 86.35 & 7.92 & \textbf{5.06} & \textbf{3.71} & 5.56 \\ \cline{2-10} 
 & \textbf{20\%} & 87.08 & 82.47 & 78.63 & 82.73 & 7.44 & 6.19 & 4.42 & 6.02 \\ \cline{2-10} 
 & \textbf{30\%} & 86.80 & 82.94 & 76.87 & 82.21 & 7.05 & 5.35 & 4.78 & 5.73 \\ \hline
\multirow{4}{*}{fold 5} & \textbf{5\%} & 91.15 & 84.96 & 83.13 & 85.41 & 6.78 & 6.12 & 4.91 & 5.94 \\ \cline{2-10} 
 & \textbf{10\%} & 91.22 & 85.50 & \textbf{83.56} & 86.76 & 6.16 & 5.65 & 4.60 & \textbf{5.47} \\ \cline{2-10} 
 & \textbf{20\% }& \textbf{91.29} & 85.44 & 82.58 & 86.44 & 7.70 & 6.72 & 5.82 & 6.74 \\ \cline{2-10} 
 & \textbf{30\% }& 90.70 & 84.70 & 82.53 & 85.97 & 6.82 & 5.87 & 4.76 & 5.82 \\ \hline
\end{tabular}
\caption{Validation performance metrics comparison across data augmentation probabilities and folds applied to Optimized U-Net model. Metrics include Dice and HD95 for
whole tumor (WT), tumor core (TC), and enhancing tissue (ET).
}
\label{tab:performance}
\end{table}

\subsection{BraTS-Africa}
For the test phase of the BraTS-Africa challenge, we submitted 4 results, namely: the 3D adaption of RhizoNet \cite{Sordo2024} trained on the BraTS-Africa dataset, the Optimized U-Net architecture trained on both the BraTS-Africa dataset and the BraTS 2023 Adult Glioma dataset (shown as \enquote{Combined Dataset} in \ref{table:unet_metrics_SSA}), and the DANN architecture trained on both the BraTS-Africa dataset and the BraTS 2023 Adult Glioma dataset. Additionally, we submitted a result of an ensemble of the the predictions from the DANN model, the DANN model without gradient reversal, and the Combined Dataset training, shown as \enquote{DAEnsemble} in \ref{table:unet_metrics_SSA}.

\begin{table}[H]
\centering
\resizebox{\linewidth}{!}{
\begin{tabular}{|c|ccc|ccc|ccc|ccc|}
\hline
\multirow{2}{*}{Model Architecture} & \multicolumn{3}{c|}{Dice [\%]} & \multicolumn{3}{c|}{HD95 [mm]} & \multicolumn{3}{c|}{LW Dice [\%]} & \multicolumn{3}{c|}{LW HD95 [mm]} \\
\cline{2-13}
& WT & ET & TC & WT & ET & TC & WT & ET & TC & WT & ET & TC \\
\hline
RhizoNet & 83.46& \textbf{81.76}& \textbf{83.27}& 33.16& 25.78& 27.58& 49.07& 66.72& 67.25& 172.24& 87.11& 90.25\\
Combined Dataset &{ 94.46} & 80.87 & 80.39 & 4.23 & 14.55 & \textbf{9.10} & \textbf{93.96} & 77.86 & \textbf{76.24} & 4.30 & 26.29 & \textbf{25.58} \\
DAEnsemble & 94.42 & 75.78 & 73.86 & 4.14 & 15.85 & 20.12 & 93.76 & 72.95 & 70.20 & 4.24 & 27.60 & 36.54 \\
DANN & 94.47 & 81.68 & 77.30 & 3.97 & \textbf{14.47} & 25.10 & 93.86 & \textbf{78.86} & 71.26 & 4.06 & \textbf{26.23} & 31.90 \\
DANN w/o Grad. Reverse& \textbf{94.50} & 80.64 & 76.04 & \textbf{3.41} & 15.09 & 25.08 & 93.39 & 77.77 & 69.96 & \textbf{3.54} & 26.84 & 31.93 \\
\hline
\end{tabular}
}
\caption{Test performance metrics for the different training pipelines considered. Metrics include Dice, HD95, lesion-wise Dice (LW Dice), and lesion-wise HD95 for whole tumor (WT), tumor core (TC) and ET.}
\label{table:unet_metrics_SSA}
\end{table}


\subsection{BraTS-Inpainting}
Both Attention UNet frameworks were trained for 1500 epochs using supervised criterion and conditional GANs based criterion and converged to produce similar performance in terms of validation metrics, with GANs based method performing slightly better on validation datasets. MA3T-F with GrokFast was able to achieve comparably high validation metrics after only 20 epochs of training indicating a superior framework for inpainting tasks. Although these results are not state of the art, with more epochs, MA3T-F with GrokFast has the potential to surpass the Attention Unet models.
\begin{table}[H]
\centering
\begin{tabular}{|c|c|c|c|}
\hline
\textbf{Model} & \textbf{SSIM} & \textbf{PSNR} & \textbf{MSE} \\ \hline
AUNet(1500eps) & 0.9995 & 20.1592 & 0.0002 \\ \hline
AUNet+GANs(1500eps) & 0.9997 & 20.4588 & 0.0002 \\ \hline
MA3T-Fv1+GANs(20eps) & 0.9996 & 16.2408 & 0.0015 \\ \hline
\end{tabular}
\caption{Validation performance metrics for various frameworks tested for the inpainting task. Metrics include SSIM(closer to 1 is better), PSNR(higher is better), and MSE(smaller is better).}
\label{tab:synthesis_model_metrics}
\end{table}

\section{Discussion}
When comparing various deep learning models for segmentation tasks, it is evident that nn-UNet stands out as the superior model when evaluated individually. However, U-Mamba shows impressive performance, particularly when considering its significantly shorter runtime compared to other models.

Employing ensemble techniques can greatly boost the robustness of the models, and incorporating effective post-processing strategies is essential for accurate tumor segmentation across various scenarios. Additionally, post-processing plays a crucial role in refining model outputs, specifically in minimizing false positives, which is particularly important when evaluating the model on different types of brain tumors.

Attention UNet achieves similar metrics in both supervised and WGANs based frameworks, with WGANs being slightly better overall. This could indicate saturation of the model in terms of performance and a need for a stronger generator architecture.
MA3T-F model achieved high metric scores after only 20 epochs.
This can be attributed to the use of Gated Linear Attention mechanism which significantly reduces computational complexity and efficiently uses memory for improved latency during training and inference, further allowing for much larger models to be trained on similar hardware. The Gated Linear Attention layers are also able to better capture global contextual information for inpainting of large regions of missing tissue. This global contextual information is further reinforced by incorporation of the Fourier Blocks in parallel. 
With the incorporation of GrokFast algorithm, the MA3T-F approaches generalization much faster compared to the other experiments where GrokFast was not utilized.

For the BraTS Africa models considered, the DANN architecture produced the best overall results, with first or second place results in most metrics from validation.

\section{Acknowledgments}The work of Juampablo Heras Rivera was partially supported by the U.S. Department of Energy Computational Science Graduate Fellowship under Award Number DE-SC0024386. 
The authors are especially grateful to Dr. Udunna Anazodo and Maruf Adewole.

%
%

%
%
%
%
\bibliography{bib}

\end{document}